\documentclass[aps,pra,floats,twocolumn,showpacs,superscriptaddress]{revtex4-1}
\usepackage[dvips]{graphicx}
\usepackage{epsfig}
\usepackage{subfigure}
\usepackage{amsmath}
\usepackage{amssymb}
\usepackage{epsfig}
\usepackage{hyperref}
\usepackage{color}

\begin{document}

\title{Superradiance, charge density waves and lattice gauge theory in a generalized Rabi-Hubbard chain}

\author{Axel Gagge}
\author{Jonas Larson}
\affiliation{Department of Physics, Stockholm University, Se-106 91
  Stockholm, Sweden}
  \date{\today}

\begin{abstract}
We investigate a one-dimensional Rabi-Hubbard type of model, arranged such that a qdot is sandwiched between every cavity. The role of the qdot is to transmit photons between neighboring cavities, while simultaneously acting as a photon non-linearity. We consider three-level qdots in the $\Lambda$ configuration, where the left and right leg couples exclusively to the left or right cavity. This non-commuting interaction leads to two highly entangled incompressible phases, separated by a second order quantum phase transition: the qdot degrees-of-freedom act as a dynamical lattice for the photons and a Peierls instability breaks a second $\mathbb{Z}_2$ symmetry which leads to a dimerization in entanglement and photon number. We also find a normal insulating phase and a superfluid phase that acts as a quantum many-body superradiant phase. In the superradiant phase, a $\mathbb{Z}_2$ symmetry is broken and the phase transition falls within the transverse field Ising model universality class. Finally, we show that a limit of the model can be interpreted as a $\mathbb{Z}_2$ lattice gauge theory.
\end{abstract}


\maketitle

\section{Introduction}

Light has since always served as a tool for detecting states of matter; whether it is our own eyes registering our surrounding, or photon detectors measuring the wavelengths of light from distant stars. Light can also be used to control and change the states of matter. In modern times, for example, we can cool and trap individual particles with designed laser light. At the quantum level of single photons, strongly coupled light and matter can form novel states with no counterparts in other branches in physics~\cite{ql}; new interacting quantum many-body models derive which may host exotic phases. The study of phase transitions (PTs) in light-matter systems dates back to the early days of the laser, when the onset of lasing with increasing pump power was identified as a non-equilibrium continuous PT~\cite{laserPT}. 

Predating the laser, in 1954 Dicke showed how the rate of spontaneous emission for a set of $N$ two-level atoms could be enhanced by a factor $\sqrt{N}$~\cite{dicke1}. This phenomenon, arising due to collective multi-partite interference, has been termed \emph{superradiance} and can be derived from the Dicke model describing the coupling of $N$ identical two-level systems with a single photon mode~\cite{dicke2}. In 1973, first by Hepp and Lieb~\cite{dicke3} and shortly afterwards by Wang and Hioe~\cite{dicke4}, it was demonstrated that the Dicke model supports a second-order PT from a `normal' to a `superradiant' phase as the light-matter coupling is raised above a critical value. The corresponding PT is accompanied by a spontaneous breaking of a $\mathbb{Z}_2$ symmetry. In more recent times, there has been a strong interest in realizing PTs, like the Dicke one, in quantum optical lattice systems~\cite{cmb}. This development was spurred by the increased experimental control over many-body quantum systems and in the wake of quantum simulators~\cite{qs}.

One manifestation of a quantum PT is a non-analytic behavior of the ground-state~\cite{dickeq}. For the Dicke PT one finds such a non-analyticity, and as such it has often been referred to as a quantum PT (QPT). Nevertheless, the transition is of the mean-field type, and the role of quantum fluctuations becomes irrelevant in the thermodynamic limit~\cite{jonas1}. In this respect, the normal-superradiant transition cannot be driven by quantum fluctuations, and depending on ones personal taste one may question referring to the Dicke PT as a proper QPT. 

In a `true' many-body quantum normal-superradiant PT, as the light-matter coupling $g$ is increased the system would enter a superradiant phase while quantum fluctuations remain extensive in the thermodynamic limit. This is more in the vein of paradigm quantum critical models like the transverse field Ising or Bose-Hubbard models~\cite{sachdev}. As a way to enhance the role of quantum fluctuations one may instead consider multimode Dicke models~\cite{mdicke} or cavity arrays~\cite{jch}. In the latter, arrays of cavities/resonators are manufactured on microchips such that photons can tunnel between neighboring cavities. Each transmission line resonator is equipped with a quantum dot that acts as an artificial two-level atom~\cite{wall}. This produces a Jaynes-Cummings nonlinearity~\cite{photblock}, which acts as an effective photon-photon interaction. The resulting physics (neglecting photon losses) is essentially the same as for the Bose-Hubbard model with insulating and superfluid phases of polaritons~\cite{rwabh}. 

In a more recent work, the light-matter terms of the model were used to produce an effective interaction as well as the kinematics~\cite{gr}. The authors studied a one-dimensional array of resonators, with a two-level system (qubit) placed between neighboring resonators, such that photon tunneling is mediated by a Jaynes-Cummings interaction. It was demonstrated that the low energy physics is described by the transverse field Ising model. While not discussed in~\cite{gr}, we may think of the emerging Ising transition as a normal-superradiant PT that is driven by quantum fluctuations; in the normal phase, all qubits are to a good approximation in their lower state and the cavity modes in vacuum, while if the light-matter coupling is increased beyond a critical value $g_c$, the photon modes as well as the excited qubit states are populated.

In this work we also consider a cavity array with mediated photon tunneling, but instead of qubits we take three-level systems, i.e. qutrits with three internal states $|1\rangle$, $|2\rangle$, and $|3\rangle$. We assume strong light-matter couplings $g \sim 1$. The novel features of our model are due to the non-commuting structure of the interaction terms; every other resonator couples to the $|1\rangle\leftrightarrow|3\rangle$ transitions of its left and right qutrit, while every other instead addresses the $|2\rangle\leftrightarrow|3\rangle$ transitions~\cite{strint}.

It may be helpful to picture the photons as `site variables' and the qutrits as `bond variables', forming a `dynamical lattice' for the photons as in quantum link models~\cite{qlm}. A salient result that may arise from a dynamical lattice is an altered periodicity, leading to novel phases such as charge density waves~\cite{cdw} first predicted by Peierls in one dimension, and supersolids~\cite{ss}. Indeed, for not too large coupling strengths we find a phase (CDW) with a charge density wave order. Contrary to a traditional Peierls instability for free fermions, the filling in our model is not set by an external chemical potential but is determined mainly by the light-matter coupling $g$, and there are no strict constraints on the filling for the appearance of the CDW. This is akin the `photonic Peierls transition' recently discussed in an extended Bose-Hubbard model with a similar dynamical lattice as the one considered here~\cite{mbh}. Beyond the CDW, as the coupling is increased further, there is a phase transition to another phase (NE), characterized by unbroken translational symmetry and high pairwise entanglement over the bonds.

Neither of these phases are superradiant, i.e. polaronic superfluids. In order to open up for the aforementioned quantum normal-superradiant PT we assume a static dipolar field driving the $|1\rangle\leftrightarrow|2\rangle$ transition. Such a field has a `resetting' effect on the qutrits; a qutrit making the transition of say $|1\rangle\rightarrow|3\rangle\rightarrow|2\rangle$ by transferring one photon between two resonators can be reset by the field to its original state $|1\rangle$ without involving further photons from the resonators. Above some critical drive amplitude $s_c$ and coupling strength $g_c$ we find a superradiant phase which spontaneously breaks the $\mathbb{Z}_2$ symmetry.  For weaker couplings, on the other hand, a symmetric normal phase emerges. In addition, our numerical results indicate a multi-critical point where all four phases meet. All transitions are found to fall within the universality class of the transverse field Ising model, i.e. not of the mean-field type but truly quantum by nature.

The outline of the paper is as follows. In the next section we introduce our model and find its symmetries. In Sec.~\ref{sec3} we present the zero temperature phase diagram as obtained from DMRG calculations, and characterize the different phases. We also discuss the possible transitions between the phases. Finally, in Sec.~\ref{sec4} we conclude with a summary.

\section{Model system and analysis}\label{sec2}

\begin{figure}
\centerline{\includegraphics[width=10cm]{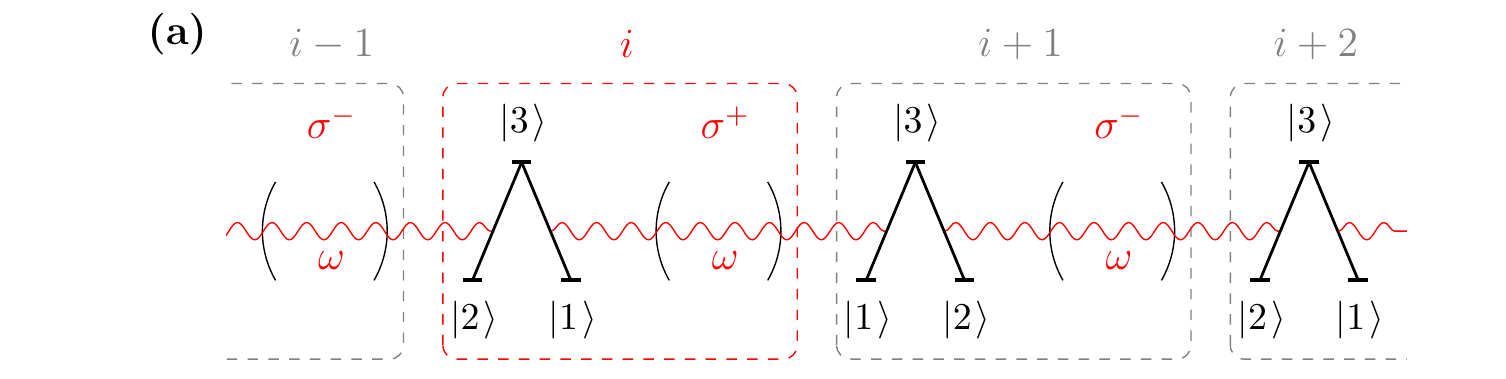}}
\centerline{\includegraphics[width=3cm]{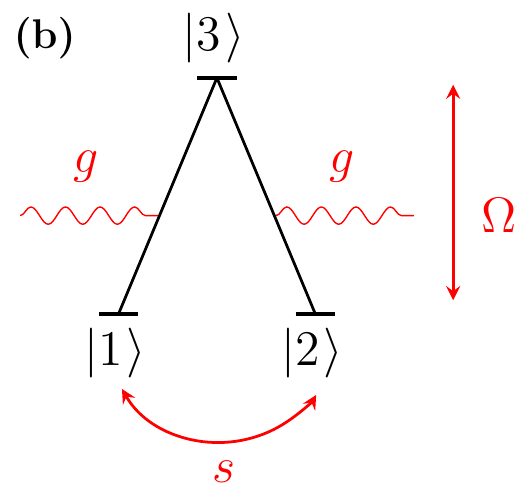}}
\caption{(Color online)  Schematic picture of the system set-up (a) and the qutrit-cavity coupling (b). An array of identical cavities supporting a single mode, with frequency $\omega$, are coupled via three-level $\Lambda$-systems on the bonds between the cavities (a). These qutrit $\Lambda$-systems host two degenerate bare ground states $|1\rangle$ and $|2\rangle$, and an excited state $|3\rangle$ separated from the lower states by $\Omega$, see (b). One of the legs of the $\Lambda$-system is coupled to the cavity to the left and the other leg is coupled to the cavity to the right. Selecting the transitions in this particular way can be achieved by alternating the polarization, $\sigma^\pm$, between the cavities. Photon transport between consecutive cavities are only accompanied by transitions within the corresponding $\Lambda$-system. In the rotating wave approximation we would have a qutrit transition $|1\rangle\rightarrow|2\rangle$ for a photon moving to the right from cavity $i$ to cavity $i+1$, and vice versa for the other direction. The counter rotating terms, neglected within the rotating wave approximation, break this property. In addition, we add a direct coupling between the $|1\rangle$ and $|2\rangle$ states with an amplitude $s$, as depicted in (b). As explained in the main text, a polaronic superfluid (superradiant phase) is only possible for a non-zero coupling $s$.}
\label{fig1}
\end{figure}

We investigate a one-dimensional array of cavities, where photon tunneling between each pair of neighbor cavities is mediated by quantum dots. The system is schematically pictured in Fig.~\ref{fig1} (a). The idea of interconnecting cavities via superconducting qubits dates back to the works of~\cite{rescon}, where the qubits were considered as controllable switches for the tunneling of photons between the cavities. Based on these ideas, it was possible to experimentally connect three resonators with the help of qubits~\cite{rescon2}. The extension to arrays of multiple cavities was also later considered~\cite{gr}. The system is then truly in the quantum many-body regime and the low energy physics can be seen as bosons living on a `dynamical lattice'~\cite{mbh}. In this work we extend the model of~\cite{gr} to three-level quantum dots, where the transitions between the qdot levels can be addressed separately. 

\subsection{Full Hamiltonian}

We assume ideal qutrits (three-level systems) in the $\Lambda$ configuration: the two states $|1\rangle$ and $|2\rangle$ are degenerate, while the state $|3\rangle$ is of a higher energy. Angular momentum must be preserved, so if, e.g., the lower states are angular momentum states with $m=\pm1$ and the excited state has $m=0$, then the qutrit transition $3 \rightarrow 1$ ($3 \rightarrow 2$) emits a $\sigma^+$-polarized ($\sigma^-$-polarized) photon. One possible method of realizing our model would be by using polarization-selective cavities. Due to these selection rules, the qutrit transition $3 \rightarrow 1$ ($3 \rightarrow 2$) can only emit a photon into the left (right) cavity. We return to the question about experimental realization in the conclusion. As mentioned in the introduction, we also introduce a static dipolar field between the two lower qutrit states $|1\rangle$ and $|2\rangle$. Note that such a field breaks the angular momentum conservation, which is possible to realize by external two photon driving~\cite{sprint2}. The level configuration and system parameters are shown in Fig.~\ref{fig1} (b). 

The full model system, as shown in Fig.~\ref{fig1} (a), consists of an infinite chain of cavities interspersed with qutrits. We label each qutrit and the resonator to its right by the same site index $i$. Cavities with even (odd) index select $\sigma^+$-polarized ($\sigma^-$-polarized) modes and are coupled to the $3 \leftrightarrow 1$ ($3 \leftrightarrow 2$) transition with a Rabi interaction of strength $g$. We do not restrict the analysis to moderate couplings $g\ll\omega$, in which the rotating wave approximation is applicable, but allow for $g\sim\omega$, i.e. the deep strong coupling regime~\cite{deep}. 

The Hamiltonian can be split in a bare and an interaction part
\begin{equation}\label{fullh}
 H= H_\mathrm{B}+ H_\mathrm{int}.
\end{equation}
The bare Hamiltonian is comprised of the harmonic oscillators representing the single light modes of the resonators, the bare energies of the qutrits, as well as the dipolar field on the two lower qutrit states. Expressed in terms of Gell-Mann matrices~\cite{gmm} (see Appendix~\ref{appGM} for the definition of the Gell-Mann matrices $\lambda^{(\alpha)}_i$), the bare Hamiltonian reads ($\hbar=1$ throughout)
\begin{equation}\label{bareh}
 H_\mathrm{B}=\omega \sum_i  a_i^\dagger a_i - \frac{\Omega}{\sqrt{3}} \sum_i \lambda^{(8)}_i-s \sum_i \lambda^{(1)}_i,
\end{equation}
where $\omega$ is the resonant angular frequency of the cavities, $\Omega$ is the frequency (energy) of the upper qutrit states, $s$ is the strength of the dipolar field on the lower qutrit states and $ a_i^\dagger$ ($ a_i$) are the photon creation (annihilation) operators for the $i$'th cavity, i.e. for a photon Fock state of the $i$'th cavity, $ a_i|n\rangle_i=\sqrt{n_i}|n-1\rangle_i$, $ a_i^\dagger|n\rangle_i=\sqrt{n_i+1}|n+1\rangle_i$, and for the number operator $ n_i|n\rangle_i= a_i^\dagger a_i|n\rangle_i=n_i|n\rangle_i$. The interaction Hamiltonian can be written as
\begin{align}\label{intham}
 H_\mathrm{int}=g\sum_{i \text{ odd}}\left( a_i+ a_i^\dagger\right) \left( \lambda^{(4)}_{i-1} +\lambda^{(4)}_{i}\right)  \nonumber \\
+g \sum_{i \text{ even}}\left( a_i+ a_i^\dagger\right) \left( \lambda^{(6)}_{i-1}+\lambda^{(6)}_{i} \right) 
\end{align}
where $g$ is the coupling strength.

We note that there is a duality which maps the Hamiltonian ${H}(s, g, \omega, \Omega) \rightarrow {H}(-s, g, \omega, \Omega)$. The unitary transformation is
\begin{align}\label{duality}
{U} = \exp\left( i \pi \sum_i \left( |1\rangle_{\!i\,i\!} \langle1| +  \frac{1-(-1)^i}{2} {n}_i \right) \right)
\end{align}
Hence, it is sufficient to consider $s > 0$, and we remark that for any superradiant phase for $s < 0$ there exists a corresponding superradiant phase for $s > 0$, but with a staggered ordered parameter $\langle {a}_i \rangle \propto (-1)^i$. 

\subsection{Effective qutrit Hamiltonian}

It is possible to eliminate the photon degrees-of-freedom to derive an effective model of interacting $SU(3)$ spins in one dimension. To obtain such an effective model, we employ the polaron transformation (also called Lang-Firsov transformation)~\cite{polar,lf}, which is an ansatz for the ground state relying on a time-scale separation between fast and slow variables. The method has proven efficient in describing phonon physics, where the states of the ions in a crystal are `displaced' according to the electronic state; phonons (fast variables) dress the electrons (slow variables) to form a polaronic excitation~\cite{polaron}. If the phonons are displaced, the idea is to find the new minima, i.e. the displaced phononic vacuum~\cite{polar}. The variational ansatz of the polaron transformation for our model is given by
\begin{equation}
| \Psi(\left\{ c_i \right\}, \left\{ \alpha_i \right\}, \gamma) \rangle =
U_\gamma^\dagger | \psi_\text{qutrit}(\left\{ c_i \right\}) \rangle \otimes | \vec{\alpha} \rangle,
\end{equation}
where we have defined $| \vec{\alpha} \rangle = \bigotimes_i | \alpha_i \rangle$. Here, $\left\{ c_i \right\}$ are the $3^N$ coefficients of the entire many-body state of the qutrits, $\alpha_i \in \mathbb{R}$ is the amplitude of a photonic coherent state $| \alpha_i \rangle$, and $\gamma$ is a real parameter which is later chosen such that the photonic and qutrit variables are decoupled. While this ansatz is a mean-field ansatz for the photonic variables, it allows for general quantum correlations in the qutrit variables. We define the unitary polaron transformation as
\begin{equation}
U_\gamma = \exp\left[\gamma \sum_i \left( a_i-a_i^\dagger \right) P_{i-1,i} \right],
\end{equation}
where we have defined
\begin{equation}
P_{i-1,i}=\begin{cases}
\lambda^{(4)}_{i-1} +\lambda^{(4)}_{i},\,\,\,\,\,\,\, \text{ $i$ odd,} \\
\lambda^{(6)}_{i-1}+\lambda^{(6)}_{i},\,\,\,\,\,\,\, \text{ $i$ even.} 
\end{cases}
\end{equation}
Taking the expectation value with respect to the photonic variables, $\langle \vec{\alpha}| H | \vec{\alpha} \rangle$, yields the effective qutrit Hamiltonian
\begin{equation}\label{qham}
\begin{array}{lll}
H_\text{eff} & = &  
\displaystyle{-\tilde{s} \sum_i \lambda^{(1)}_i - \frac{\tilde{\Omega}}{\sqrt{3}} \sum_i \lambda^{(8)}_i-J \sum_\text{$i$ odd} \lambda^{(4)}_{i-1} \lambda^{(4)}_{i}}\\ \\
& & \displaystyle{ - J \sum_\text{$i$ even} \lambda^{(6)}_{i-1} \lambda^{(6)}_{i},}
\end{array}
\end{equation}
where $J$, $\tilde{s}$, and $\tilde{\Omega}$ are renormalized couplings. A detailed calculation of Eq.~(\ref{qham}), as well as the expressions for the renormalized couplings, are given in the Appendix~\ref{appendix}.

\subsection{The normal-superradiant phase transition}

As motivated in the introduction, one reason for studying the present model is to find a  truly quantum normal-superradiant PT. Similar to the Dicke model but with extensive quantum fluctuations in the thermodynamic limit. Our Hamiltonian is invariant under a $\pi$-rotation with respect to the total excitation number
\begin{equation}
N_\mathrm{ex}=\sum_i\left(  n_i+\sqrt{3}\lambda_i^8 \right),
\end{equation}
i.e. the corresponding unitary is
\begin{equation}
\Pi=\exp(-i \pi N_\mathrm{ex})=(-1)^{ N_\mathrm{ex}},
\end{equation}
Since the eigenvalues of $ N_\mathrm{ex}$ are integers, we clearly have $\Pi^2=\mathbb{I}$ as required for a $\mathbb{Z}_2$ symmetry. That the Hamiltonian is symmetric under the action of $ \Pi$ follows from noticing
\begin{equation}\label{pari}
\begin{array}{lll}
\Pi a_i \Pi^\dagger=- a_i, & & \Pi \lambda^{(4)}_i  \Pi^\dagger=-\lambda^{(4)}_i, \\ \\
\Pi \lambda^{(6)}_i  \Pi^\dagger=-\lambda^{(6)}_i, & & \Pi \lambda^{(8)}_i  \Pi^\dagger= \lambda^{(8)}_i.
\end{array}
\end{equation} 

Apart from a few special exceptions, continuous phase transitions occur when the ground state of a system spontaneously breaks the symmetries of its Hamiltonian~\cite{GL}. We can then find a (local) order parameter, which is zero in the symmetric phase and non-zero in the symmetry broken phase. For the normal-superradiant PT, which is connected to the breaking of a similar $\mathbb{Z}_2$ symmetry, a proper choice is $\phi=\langle a\rangle$~\cite{dickeexp}. Given a real light-matter coupling $g$, the order parameter will be real; the sign of $\phi$ determines the parity~\cite{jonas1}. Similarly in our model, breaking of the parity symmetry results in a real non-zero value of
\begin{align}\label{order1}
\phi = \frac{1}{L} \sum_i \text{sign}(s)^{i} \langle a_i\rangle,
\end{align}
where $\text{sign}(s)^{i}$ ensures that the order parameter works whether the parity breaks ferro-magnetically (for $s > 0$) or anti-ferromagnetically (for $s < 0$). The magnitude of $\phi$ is the same for every lattice site, but the sign can alternate between neighboring sites; constant signs represent ferromagnetic order, while alternating signs anti-ferromagnetic order. Since the symmetry $\Pi$ involves both qutrits and resonators, there are two other possible order parameters which should show the same (anti)-ferromagnetic order:
\begin{equation}\label{order1b}
\phi_4 = \frac{1}{L} \sum_i \text{sign}(s)^{i}\langle \lambda^{(4)}_i \rangle, 
\end{equation}
\begin{equation}
\phi_6 = \frac{1}{L} \sum_i \text{sign}(s)^{i} \langle \lambda^{(6)}_i \rangle.
\end{equation}

As a one-dimensional spin chain supporting a $\mathbb{Z}_2$ symmetry, the critical point should belong to the universality class of the transverse field Ising model. To elucidate this, we investigate the limit of $s \to \infty$ and extrapolate to finite $s$. By a Hadamard rotation of the qutrit states
\begin{equation}
\begin{array}{l}
| 1 \rangle_i \to \left( | 1 \rangle_i + | 2 \rangle_i \right) / \sqrt{2}, \\
| 2 \rangle_i \to \left( | 1 \rangle_i - | 2 \rangle_i \right) / \sqrt{2}, \\
| 3 \rangle_i \to | 3 \rangle_i,
\end{array}
\end{equation}
the Hamiltonian can be rewritten as
\begin{equation}
\begin{array}{lll}
 H'_\mathrm{B} & = & \displaystyle{\omega\sum_i a_i^\dagger a_i-\Omega \sum_i \lambda^{(8)}-s\sum_i \lambda^{(3)},}\\ \\
 H'_\mathrm{int} & = & \displaystyle{g \sum_{i \text{ odd}} \left( a_i+ a_i^\dagger\right)\left( \frac{\lambda^{(4)}_{i-1} + \lambda^{(6)}_{i-1}}{\sqrt{2}} + \frac{\lambda^{(4)}_i + \lambda^{(6)}_i}{\sqrt{2}} \right)}\\ \\
& & \displaystyle{+g \sum_{i \text{ even}} \left( a_i+ a_i^\dagger\right)\left( \frac{\lambda^{(4)}_{i-1} - \lambda^{(6)}_{i-1}}{\sqrt{2}} + \frac{\lambda^{(4)}_i - \lambda^{(6)}_i}{\sqrt{2}} \right).}
\end{array}
\end{equation}
The bare Hamiltonian has eigenvalues
\begin{equation}
E_{1n} = s + \omega n,\,\,\,\,E_{2n} = -s + \omega n,\,\,\,\,E_{3n} = \Omega + n.
\end{equation}
The population in the state with bare energy $E_{2n}$ will be small if $s$ is large enough and the mean photon number $\bar{n}$ of each resonator is small. In this limit, we project the qutrit degrees of freedom of the Hamiltonian by $P = \prod_i \left( |1\rangle_{\!i\,i\!} \langle1| + |3\rangle_{\!i\,i\!} \langle3| \right)$ to find
\begin{equation}
\begin{array}{lll}
{P}  H'_\mathrm{B} {P} & = & \displaystyle{\omega\sum_i a_i^\dagger a_i-\frac{\sqrt{3} \Omega}{2} \sum_i \sigma^z_i+ \text{const.},} \\ \\
{P}  H'_\mathrm{int} {P} & = &  \displaystyle{\frac{g}{\sqrt{2}} \sum_i \left( a_i+ a_i^\dagger\right)\left( \sigma^x_i + \sigma^x_{i-1} \right),}
\end{array}
\end{equation}
where $\sigma^z_i = |1\rangle_{\!i\,i\!}\langle1| - |3\rangle_{\!i\,i\!}\langle3|$ and $\sigma^x_i = |1\rangle_{\!i\,i\!}\langle3| + |3\rangle_{\!i\,i\!}\langle1|$. In this limit, the model is equivalent to that considered by~\cite{gr}, who already concluded that the phase transition falls within the Ising universality class. By the duality~(\ref{duality}), we conclude that the Hamiltonian is `anti-ferromagnetic' for $s < 0$.

\subsection{Quasi-translation symmetry}

Assuming periodic boundary conditions, there is a second global symmetry of the model which we call \emph{quasi-translation symmetry}: a permutation of the qutrit states $|1\rangle, |2\rangle$, followed by a translation with one site. Calling it $\tilde{\mathcal{T}}$, the action on the operators is
\begin{equation}
\begin{array}{lll}
\tilde{\mathcal{T}} a_i \tilde{\mathcal{T}}^\dagger = a_{i+1}, & & \tilde{\mathcal{T}} \lambda^{(3)}_i \tilde{\mathcal{T}}^\dagger = -\lambda^{(3)}_{i+1}, \\ \\
\tilde{\mathcal{T}} \lambda^{(4)}_i \tilde{\mathcal{T}}^\dagger = \lambda^{(6)}_{i+1}, & & \tilde{\mathcal{T}} \lambda^{(6)}_i \tilde{\mathcal{T}}^\dagger = \lambda^{(4)}_{i+1}, \\ \\
\tilde{\mathcal{T}} \lambda^{(\alpha)}_i \tilde{\mathcal{T}}^\dagger = \lambda^{(\alpha)}_{i+1}, & & \\
\end{array}
\end{equation}

for $\alpha = 1, 8$. Since this symmetry involves no change in parity of the photonic operators, no superradiance is involved in a corresponding symmetry breaking phase transition. It can still be called a parity symmetry, since $\tilde{\mathcal{T}}^2 = 1$, but the parity change involves the qutrit operator $\lambda^{(3)}_i$ rather than $\lambda^{(4)}_i$ and $\lambda^{(6)}_i$ as in Eq.~(\ref{pari}). For the spontaneous breaking of the quasi-translation symmetry $\tilde{\mathcal{T}}$, we introduce the order parameter
\begin{equation}\label{order2}
\varphi=\frac{1}{L} \sum_i \langle \lambda^{(3)}_i \rangle.
\end{equation}

\subsection{Local gauge symmetries for $s=0$}
In the limit that $s=0$, we find an infinite set of \emph{local} conserved quantities
\begin{equation}\label{locsym}
\Pi_i=\begin{cases}
B^1_i Q_i B^1_{i+1} \,\,\,\,\,\,\,\,\,\,\,\,\,\,\,\, \text{ $i$ odd} \\
B^2_i Q_i B^2_{i+1} \,\,\,\,\,\,\,\,\,\,\,\,\,\,\,\, \text{ $i$ even,}
\end{cases}
\end{equation}
where $B^k_i = e^{i \pi | k \rangle_{\!i\,i\!} \langle k | } $ ($k=1,\,2$). We note that the global parity symmetry $\Pi = \bigotimes_i \Pi_i$ and that $\Pi_i$ have the form of generators of a $\mathbb{Z}_2$ gauge theory, where the \emph{photons} play the role of the `matter field' and the qutrits are bond variables~\cite{qlm}. Gauge theories can lead to interesting phenomena such as confinement, but in one-dimensional quantum systems the gauge theory is `trivial'~\cite{kogut}. We discuss further aspects of gauge theory in the conclusion.

It is unclear whether the presence of these gauge symmetries implies that the model is integrable along the $s=0$ line. For the quantum Rabi model, comprising a single spin-$1/2$ particle interacting with one boson mode, the $\mathbb{Z}_2$ symmetry implies that the model is integrable~\cite{braak}. The present model cannot, however, be thought of as interconnected Rabi models since we rather have an $SU(3)$ algebra instead of an $SU(2)$ one. 

\section{Phase diagram}\label{sec3}

In this section we present the numerical results obtained from DMRG simulations using the TenPy library~\cite{dmrg}. In principle, for gapped systems, DMRG will be able to reproduce the zero temperature phase diagram to arbitrary accuracy provided one keeps large enough \emph{bond dimension}~\cite{dmrg2}. In practice, to extract the critical exponents can be computationally costly. For bosons, DMRG relies on truncating the infinite local Hilbert space dimension on every site. In our model, to reach numerical convergence close to the critical point may imply including up to 10 Fock states per site, making the simulations rather time-consuming. To circumvent this slow convergence, we perform some of the DMRG simulations for the effective qutrit model~(\ref{qham}) which shares the symmetries and qualitative features of the full model. This is justified when we focus on universal critical features, and not quantitative results. 

\begin{figure}
\centerline{\includegraphics[width=8cm]{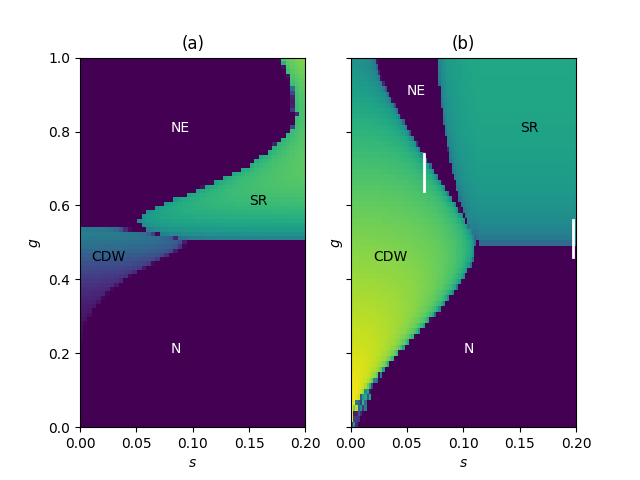}}
\caption{(Color online) The sum of the order parameters $\phi_4 + \varphi$ defined in equations~(\ref{order1b}) and~(\ref{order2}), calculated with DMRG for the full boson-qutrit model (a) and the effective qutrit model (b). In the simulations, the maximal bond dimension $\chi = 100$ was used and the bosonic Hilbert spaces were truncated to dimension 4. On the two axes we have $s$ and $g$. For larger values of the light-matter coupling $g$, the photon number $\langle n_i\rangle$ increases and we expect a breakdown of the adiabatic elimination. This is seen by the quantitative differences between the phase diagrams. Nevertheless, the structure of the phases are the same. This supports the claim that the two models share identical universal properties. We distinguish between four different phases; normal (N), charge density wave (CDW), superradiant (SR), and normal-entangled (NE). The remaining (dimensionless) parameters used for the calculation are $\omega=1$ and $\Omega=1$. In figure (b), the sweeps made to calculate figures~\ref{fig4} and \ref{fig5} are indicated with white lines.}
\label{fig2}
\end{figure}

In Fig.~\ref{fig2} we show the sum of the two order parameters, (\ref{order1b}) and (\ref{order2}), for the full model and the effective qutrit model in the $(s, g)$-plane. The characteristics of the four phases, normal (N), charge density wave (CDW), superradiant (SR), and normal-entangled (NE), will be specified below. We were not able to determine conclusively whether the four phases merge in a single multi-critical point involving all four phases, or in two tri-critical points involving only three phases each. Nevertheless, within the precision of the numerics, we lean towards the former scenario. In obtaining the figures, we used a maximal bond dimension of $\chi=100$. It was checked that setting a higher bond dimension did not improve the quality of the figure. We remark that, while the qualitative structure of the phase diagrams are the same, the polaron ansatz seems to give a bad quantitate approximation in most of the phase diagram. This is in contrast to~(\cite{gr}) who find quantitative agreement with a polaron ansatz.

There are two symmetry-breaking phases in the model: $\phi \neq 0$ in the SR phase, and $\varphi\neq0$ in the CDW phase. We mentioned above that the SR phase can alternatively be viewed as a superfluid, in contrast to the CDW phase which is an insulator in the sense that the superfluid order parameter $\phi=0$. We find no phase with both symmetries broken simultaneously, i.e. a supersolid. The N and NE phases are both symmetric with respect to both symmetries.

\begin{figure}
\centerline{\includegraphics[width=8cm]{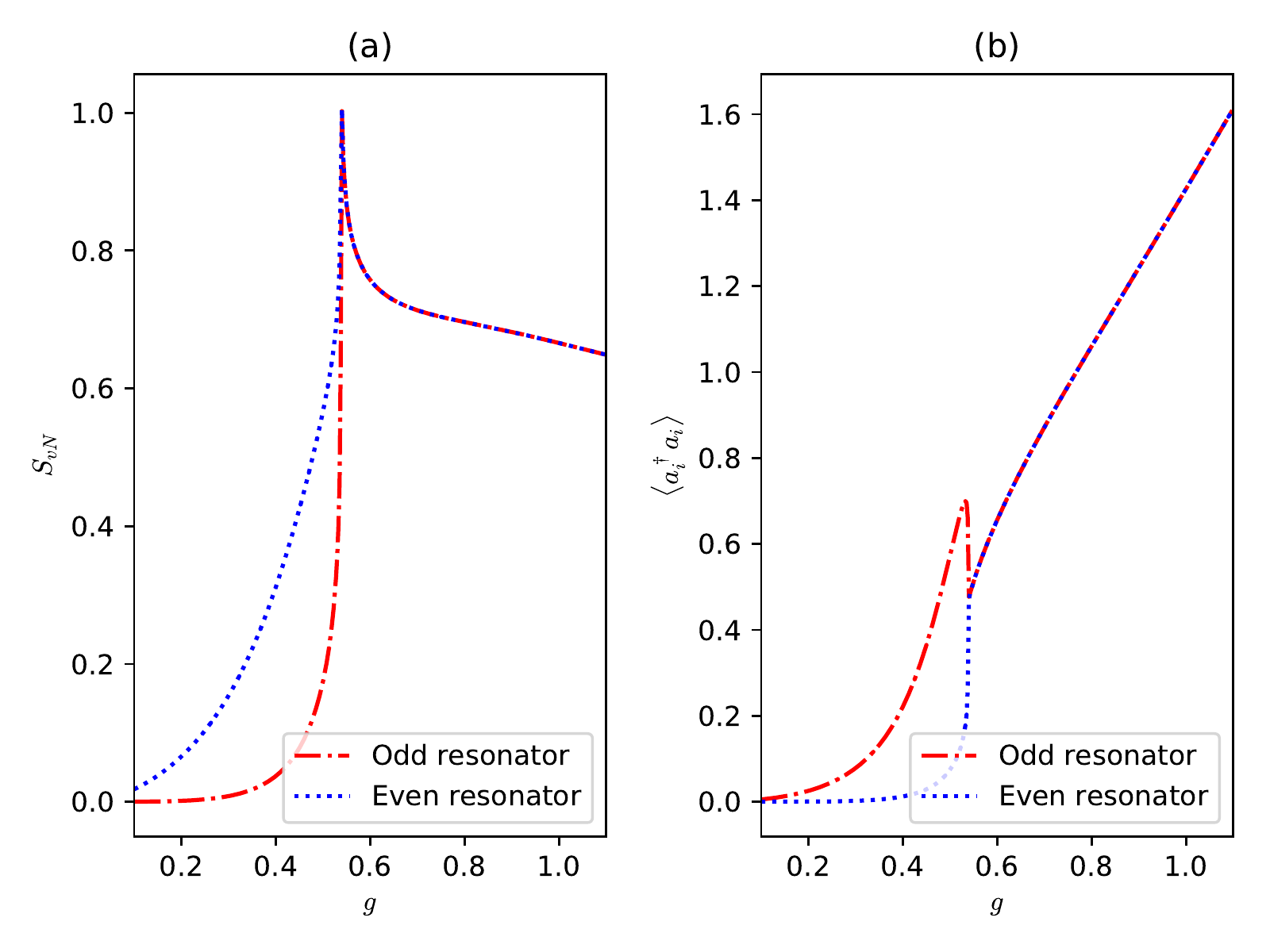}}
\caption{(Color online) (a) Entanglement entropy across both the bonds between resonators and qutrits, calculated with DMRG for the full boson-qutrit model with $s = 0, \Omega = \omega = 1$ and a bond dimension $\chi = 100$. The bosonic Hilbert spaces were truncated to dimension 4. The entanglement entropy is identical between resonators of odd index and the left and right adjacent qutrits, but differs from the entanglement between resonators of even index and its adjacent resonators. (b) Expectation value of the photon number operator $\langle a_i^\dagger a_i \rangle$, for odd (dashed line) and even (dotted line) indices, calculated from the same DMRG simulations. The spontaneous breaking of the $\tilde{\mathcal{T}}$ symmetry (the CDW in photon number) is evident in the "charge" dimerization in (b) and in the staggered entanglement entropy around odd and even resonators in (a). Note that entanglement entropy is high inside both phases, such that a non-trivial bond dimension $\chi$ is needed.}
\label{CDW-VBS-full}
\end{figure}

The normal (N) and the normal-entangled (NE) phase are both fully symmetric, and thus considered identical in the Landau classification of phases. One qualitative difference between them is the entanglement entropy~\cite{area}. We consider the von Neumann entropy
\begin{equation}\label{ent}
S_\mathrm{vN}=-\mathrm{Tr}_1\left[\rho_1\log\rho_1\right],
\end{equation}
where $\rho_1$ is the reduced density operator of subsystem 1, obtained by tracing the full density operator over the degrees-of-freedom of subsystem 2. It is understood that the trace in the above expression is over the degrees-of-freedom for subsystem 1. The subsystems 1 and 2 are defined from splitting the full system into two equal halves. We expect an `area law' behavior of the entropy in gapped phases, i.e. up to logarithmic corrections the entropy is proportional to the length of the boundary between the two subsystems~\cite{area}, which in one dimension is a single bond. For gapless phases, on the other hand, we anticipate a `volume law' where the entropy instead grows linearly with the subsystem size. The same linear growth in entanglement is expected at the critical points. In all phases, we find that the entanglement entropy saturates 

In Fig.~\ref{CDW-VBS-full}, we plot the expectation value of the photon number as well as the entanglement entropy across the CDW-NE phase transition for the full model of qutrits and resonators. In obtaining the figures, the matrix product states were truncated to a bond dimension of $\chi=100$ and the bosonic Hilbert spaces were truncated to dimension 4. The entanglement entropy in the two phases is shown in Fig.~\ref{CDW-VBS-full} (a); in the CDW phase, the entanglement is higher between resonators of odd/even index and its adjacent qutrits. The state chooses a higher entanglement around odd or even sites. This staggered structure of the entanglement entropy is another indication of the breaking of the $\mathbb{Z}_2$ quasi-translation symmetry. Indeed, in Fig.~\ref{CDW-VBS-full} we see that also the photon number breaks quasi-translation symmetry. The CDW is an entangled state which is best described by a MPS (matrix product state) of moderate bond dimension. Observe that the NE phase is fully symmetric. This is reflected in that entanglement is constant in all bonds, and no superradiance is observed. It is a gapped phase since the DMRG simulations converge for a moderate but quite high bond dimension. In the entanglement structure and MPS description, the NE phase is similar to the AKLT state. This paradigm many-body spin state has been found to be a symmetry-protected topological state for spin $S$ of odd integer values. One telltale sign of a symmetry-protected state is a two-fold degeneracy in the entanglement spectrum of the MPS. We have found no such degeneracy in the entanglement spectrum, which indicates that the NE phase is \emph{not} a symmetry-protected topological state~\cite{aklt2,oshikawa}. This raises one issue: if the Landau paradigm is not enough to distinguish the N phase from the NE phase, and no topological~\cite{top} feature distinguishes them, there is the possibility that the N phase could be smoothly deformed into the NE phase. 



\begin{figure}
\centerline{\includegraphics[width=8cm]{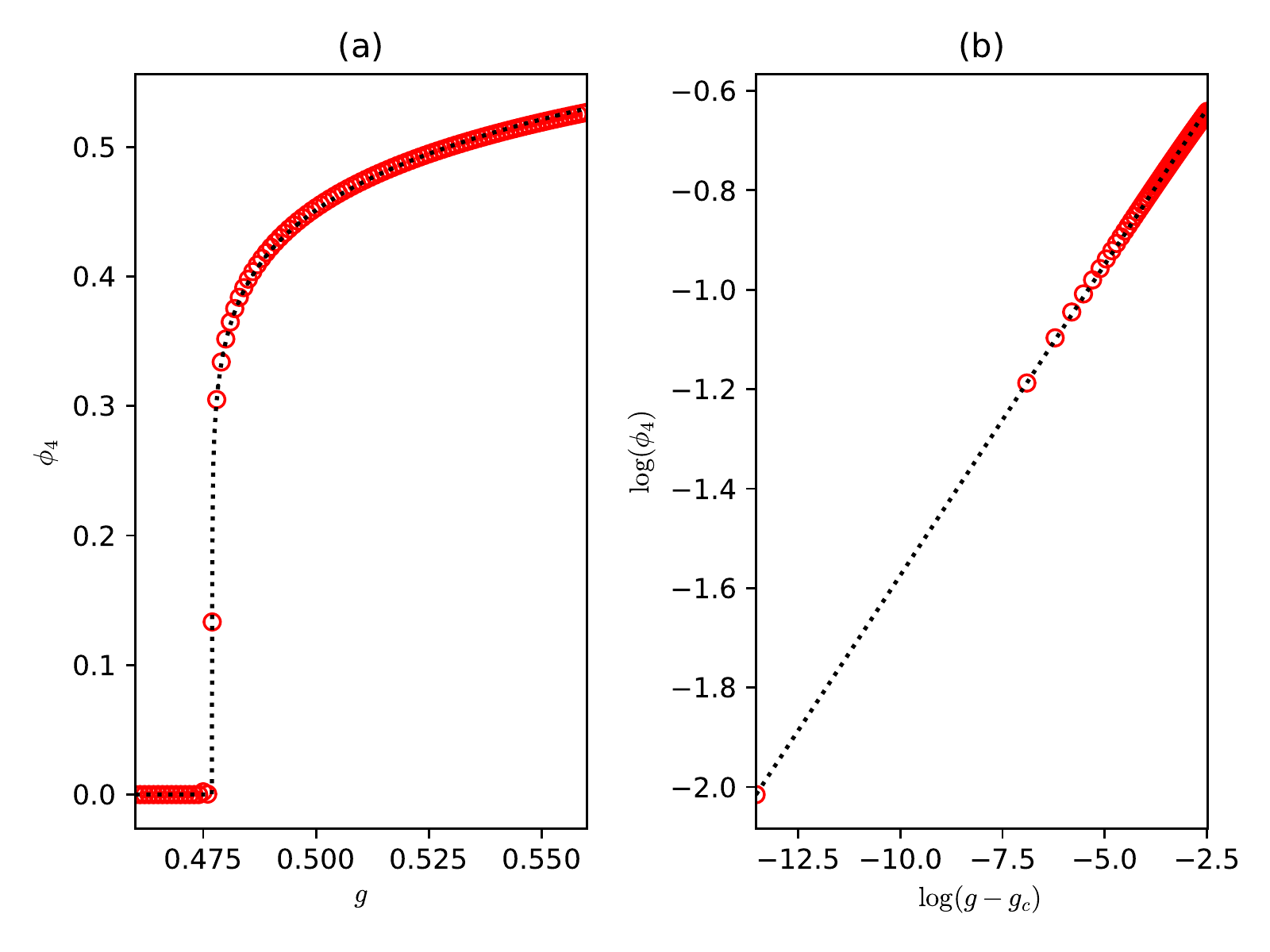}}
\caption{(Color online) Plot (a) and log-plot (b) of the order parameter~(\ref{order1b}) over the phase boundary between the N and SR phases. The circles give the numerical DMRG results calculated for the effective qutrit model~(\ref{qham}), while the solid lines are curve fits with the exponent $\beta=1/8$. The agreement is convincingly good. The remaining parameters were taken as $\omega = \Omega=1$ and  $s=0.2$.}
\label{fig4}
\end{figure}

Let us briefly return to the discussion about the non-commuting interactions. Assume no dipolar field on the qutrits, i.e. $s=0$, and assume that all qutrits are initially prepared in states $|1\rangle_i$ ($|2\rangle_i$). Then, photons are only able to tunnel between sites $i, i+1$ for $i$ odd (even), precisely because of the non-commuting structure of the interactions. This `blockade' causes the ground state to be an insulator, either of CDW or NE type (the $s=0$ line in the phase diagram). These two phases survive for small non-zero couplings $s$, but for some critical $s_c$ the two phases terminate. If the light-matter coupling $g$ is small, the photon number remains small and the system stays an insulator even for large $|s|$ values -- the normal phase. In order to find a superfluid state, $g$ has to be increased and the system enters into the SR phase. This is further evidence of a charge density wave and not the aforementioned supersolid. One could possibly imagine that the vanishing order parameters, $\phi=0$ and $\varphi=0$, in the NE phase is some numerical artifact and the two symmetries are indeed broken. However, this is physically unlikely since for at least $s=0$ we should find an insulator with $\phi=0$. 

\begin{figure}
\centerline{\includegraphics[width=8cm]{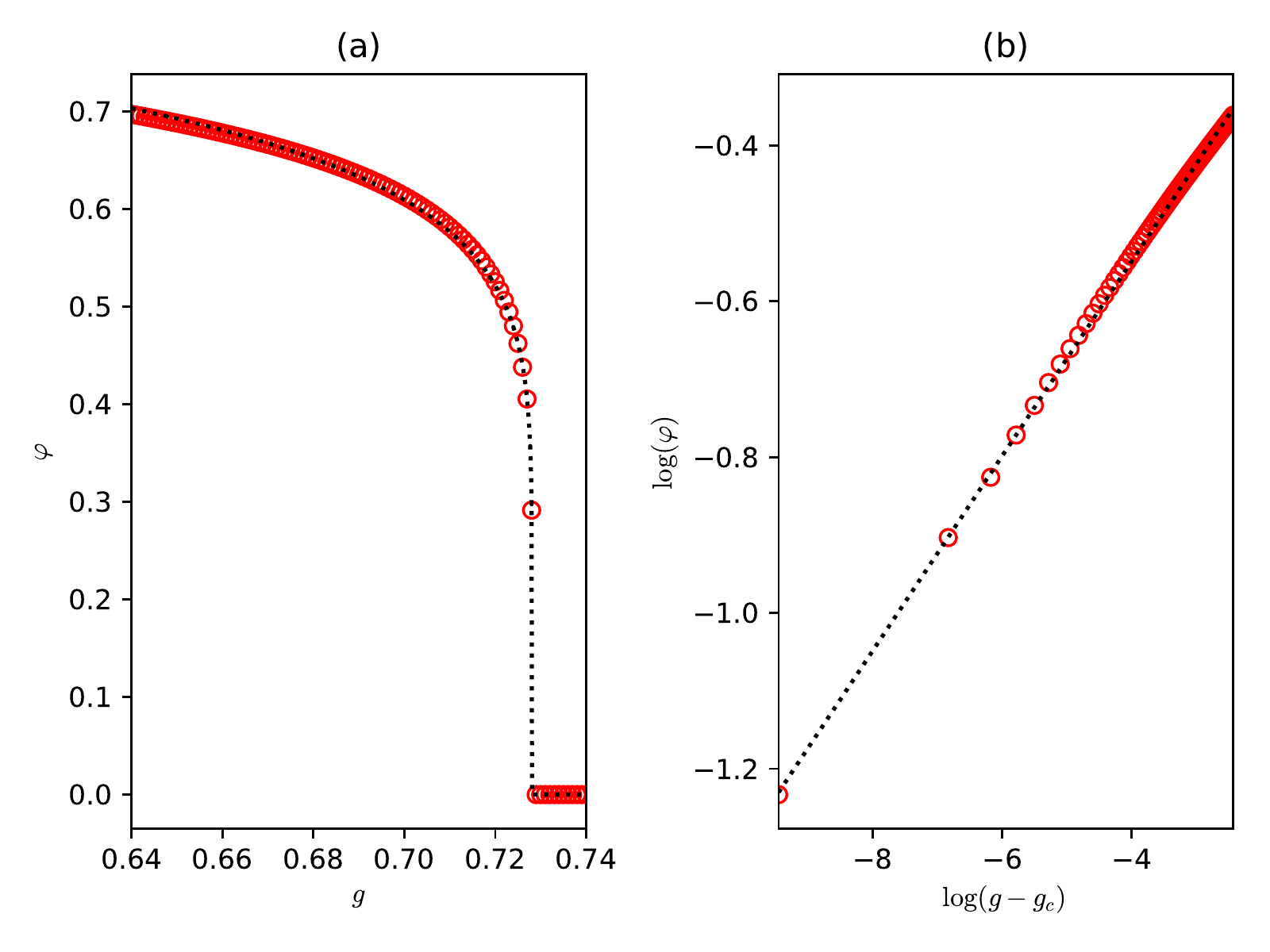}}
\caption{(Color online) The same as fig.~\ref{fig4} but for the order parameter~(\ref{order2}) across the phase boundary between the CDW and NE phases. Again we find the exponent $\beta=1/8$. The other parameters were $\omega = \Omega=1$ and  $s=0.065$.}
\label{fig5}
\end{figure}

Having identified the phases, we also wish to determine the types of transitions. With the two $\mathbb{Z}_2$ symmetries broken in the SR and CDW phases, one expects that the PTs fall within the universality class of the transverse field Ising model in one dimension~\cite{sachdev}. Indeed, we find that, close to the phase transitions, our numerically calculated order parameters can be fitted with excellent agreement to a power law with the exponent $\beta=1/8$, i.e.
\begin{equation}\label{exps}
\begin{array}{c}
\phi_4 \propto \Theta(g-g_c) (g-g_c)^{1/8}, \\ \\
\varphi \propto \Theta(s-s_c) (s-s_c)^{1/8},
\end{array}
\end{equation}
where $\Theta(x)$ is the Heaviside step function. This is demonstrated in figs.~\ref{fig4} and \ref{fig5} displaying the order parameters~(\ref{order1b}) and~(\ref{order2})  respectively. The numerical data are complemented with the expected fits~(\ref{exps}). We cross the critical points by varying $g$, while keeping $s$ fixed. The symmetry broken phases SR and CDW can be reached from either the N or the NE phases, and in all cases we have numerically verified that the exponent $\beta=1/8$ is obtained. In  fig.~\ref{fig4}, a maximal bond dimension of $\chi = 100$ was used, while in \ref{fig5}, the maximal bond dimension was $\chi = 200$.

\section{Summary and concluding remarks}\label{sec4}

In this work we have studied a generalized Rabi-Hubbard chain. A few aspects are crucially different between our model and that of the standard Rabi-Hubbard chain, which motivated our study. Let us summarize these below and explain how our work differs from earlier findings:

\begin{enumerate}
\item The qdots are not contained within the resonators but rather positioned between consecutive resonators. Photon tunneling between any two resonators occurs via the qdot, and as such the qdot acts as a `bond variable' like in the quantum link models that have been analyzed for dynamical gauge theories~\cite{qlm}. Thus, we can picture the system as bosons in a dynamical lattice -- the lattice itself constitutes quantum degrees-of-freedom. We know that dynamical lattices can give rise to novel phases and phenomena -- the most well-known example being the Peierls instability~\cite{peierls}. It is not restricted to fermions (electrons) coupled to lattice bosons (phonons), but may occur also for bosons in a dynamical lattice~\cite{mbh} and in spin systems~\cite{peierl2}. A similar transition was also found in our system, manifested as the appearance of a CDW phase of period 2. 

\item While the qdots are necessary for any dynamics (they couple the bare photon oscillators), they also serve as a nonlinear medium, i.e. the finite local Hilbert space dimension implies a nonlinear spectrum and the occurrence of insulating phases. 

\item Instead of qubits we considered qutrits. For qubits the physics is qualitatively described by a transverse field Ising model~\cite{gr}, with the insulating (normal) phase being the symmetric one, and the superfluid (superradiant) phase the symmetry broken one. In such a model, the more exotic CDW and NE phases do not appear. The specific coupling of our $\Lambda$ qutrits breaks down the local gauge symmetries $\Pi_i$ to a global symmetry $\Pi$, as explained in the main text. Without the direct coupling of the lower qutrit states, i.e. $s=0$, a qutrit acts as a single photon `diode'; if one photon has tunneled to the left, a second one cannot tunnel. Thus, the phases for $s=0$ are insulating and the order parameter $\phi$ of Eq.~(\ref{order1}) vanishes.

\item For $s=0$, the Hamiltonian is invariant under the infinite set of local symmetries~(\ref{locsym}). It has the structure of a $\mathbb{Z}_2$ gauge theory. The Mermin-Wagner theorem tells us that in one dimension a continuous symmetry cannot be spontaneously broken to generate long range order (superfluidity)~\cite{mw}. Elitzur's theorem~\cite{elitzur} tells us that gauge symmetries can never be broken, and only gauge invariant observables can have non-zero expectation values. One may speculate if this is related to the fact that we find no superfluidity in the NE phase. It is interesting to note that this phase extends to finite $s$, where the gauge theory description does not hold strictly. Note also that the superradiant phase with $\phi\neq0$ is not contradicting the Mermin-Wagner theorem since a discrete symmetry can be broken in one dimension at zero temperature.

\end{enumerate}

While our generalized Rabi-Hubbard model is simply extended from the traditional one, a most relevant question is whether it is also accessible experimentally. The main experimental novelty is the selected coupling between a qutrit and its resonators to the left and right. Such selection is easily achieved by adjusting the photon frequencies to become resonant with their respective qutrit transition, but has the disadvantage of leading to a staggered self-energy $\omega$ of the resonators. While the symmetry $\Pi$ connected to the N-SR phase transition survives, such a modification destroys the quasi-translation symmetry $\tilde{\mathcal{T}}$. By numerical simulation, we have confirmed that the SR phase is robust to such changes~\cite{strack}, while unfortunately the CDW and NE phases do not survive. Consequently, to access the full phase diagram we need to implement the selection rules by other means, for example employing polarizations as discussed in the main text. This is rather straightforward for Fabry-P{é}rot cavities, but technically more involved for transmission line resonators. Another experimental aspect is how to reach the deep strong coupling regime. This is typically done with the help of external Raman driving~\cite{carmichael}, such that the effective light-matter coupling $g$ is controlled by a classical field amplitude. Another option how to fullfil the specific selection of the transitions is to externally address different Raman transitions in the qutrit and so controlling how the frequencies of the drive fields hit the desired resonances. 

The largest experimental hindrance is, however, the loss of photons. The N--SR PT in the Dicke model survives photon losses, even if the universality class is altered~\cite{jonas1} (note that we consider the deep strong coupling regime such that the photon number in the ground state can be non-zero). It might well be that also our many-body N--SR PT would survive photon losses. How the properties of the CDW and NE phases are affected by photon losses is a most interesting issue. At the one hand, the CDW and NE phases are highly entangled and decoherence could demolish it, but on the other hand one could also argue that topology could stabilize in particular the excitations. Resolving this is certainly interesting but it lies outside the scope of the present work. Nevertheless, let us here assume that the experiment can be performed on time scales shorter than the typical photon life-time $\kappa^{-1}$, which can be of the order of ms in the microwave regime. This is of course experimentally challenging, but not completely unrealistic. 

We end by mentioning the idea of studying similar qdot generated photon tunneling in two dimensions, which should not be experimentally much more demanding. In higher dimensions the lattice geometry may play an important role -- one can imagine a plethora of different settings like lattices supporting flat bands, confinement, Dirac cones~\cite{2qs}, or other more exotic lattices~\cite{3qs}.

\begin{acknowledgements}
We thank Chitanya Joshi, Stefan Filipp, Iman Mahyaeh, Pil Saugmann and Themistoklis Mavrogordatos for helpful discussions. We acknowledge financial support from the Knut and Alice Wallenberg foundation (KAW) and the Swedish research council (VR).
\end{acknowledgements}

\appendix

\section{Gell-Mann matrices}\label{appGM}
Just as the Pauli matrices are generators of SU(2), the Gell-Mann matrices $\lambda_i$ ($i=1,\,2,\,...,8$) are generators of SU(3), i.e. they are Hermitian, traceless ($\mathrm{Tr}(\lambda_i)=0$) and orthogonal ($\mathrm{Tr}(\lambda_i\lambda_j)=2\delta_{ij}$)~\cite{gmm}. It should be clear that together with the identity matrix, any complex $3\times3$ matrix can be expressed as a linear combination of the $\lambda_i$ matrices. The usual representation of the Gell-Mann matrices is
\begin{equation}
\begin{array}{lll}
\lambda_1=\left[
\begin{array}{ccc}
0 & 1 & 0\\
1 & 0 & 0\\
0 & 0 & 0
\end{array}\right], & &  \lambda_2=\left[
\begin{array}{ccc}
0 & -i & 0\\
i & 0 & 0\\
0 & 0 & 0
\end{array}\right],\\ \\
\lambda_3=\left[
\begin{array}{ccc}
1 & 0 & 0\\
0 & -1 & 0\\
0 & 0 & 0
\end{array}\right], & &  \lambda_4=\left[
\begin{array}{ccc}
0 & 0 & 1\\
0 & 0 & 0\\
1 & 0 & 0
\end{array}\right],\\ \\
\lambda_5=\left[
\begin{array}{ccc}
0 & 0 & -i\\
0 & 0 & 0\\
i & 0 & 0
\end{array}\right], & &  \lambda_6=\left[
\begin{array}{ccc}
0 & 0 & 0\\
0 & 0 & 1\\
0 & 1 & 0
\end{array}\right],\\ \\
\lambda_7=\left[
\begin{array}{ccc}
0 & 0 & 0\\
0 & 0 & -i\\
0 & i & 0
\end{array}\right], & &  \lambda_8=\displaystyle{\frac{1}{\sqrt{3}}}\left[
\begin{array}{ccc}
1 & 0 & 0\\
0 & 1 & 0\\
0 & 0 & -2
\end{array}\right].
\end{array}
\end{equation}
Note that the Gell-Mann matrices can be written as products of matrices of the Spin 1 representation of $SU(2)$. However, only by using the Gell-Mann matrices can~(\ref{fullh}) be written as a quadratic Hamiltonian.

\section{Polaron ansatz}\label{appendix}

The goal of this appendix is to calculate the effective qutrit Hamiltonian~(\ref{qham}), which is found by taking the expectation value with respect to the photonic variables only:
\begin{align}
H_\text{eff} =  \langle \vec{\alpha} | H | \vec{\alpha} \rangle,
\end{align}
where, as in the main text, we have defined $| \vec{\alpha} \rangle = \bigotimes_i | \alpha_i \rangle$. It is useful to write the polaron unitary $U_\gamma$ in two ways:
\begin{equation}
\begin{array}{l}
U_\gamma = \exp(-i \gamma S), \\ \\
\displaystyle{S = \sum_i i \left( a_i-a_i^\dagger \right) P_{i-1,i}  = \sum_i S_{i,i+1},}
\end{array}
\end{equation}
where we have defined the two-site operators
\begin{equation}
P_{i-1,i}=\begin{cases}
\lambda^{(4)}_{i-1} +\lambda^{(4)}_{i}\,\,\,\,\, \text{ $i$ odd,} \\
\lambda^{(6)}_{i-1}+\lambda^{(6)}_{i}\,\,\,\,\, \text{ $i$ even,} 
\end{cases}
\end{equation}
and
\begin{equation}
\begin{array}{l}
S_{i,i+1}=\begin{cases}
-\sqrt{2} \left( p_i \lambda^{(4)}_i + \lambda^{(6)}_i p_{i+1} \right) \,\,\,\,\, \text{ $i$ odd,} \\
-\sqrt{2} \left( p_i \lambda^{(6)}_i + \lambda^{(4)}_i p_{i+1} \right) \,\,\,\,\, \text{ $i$ even,} 
\end{cases} \\ \\
\displaystyle{p_i = -\frac{i}{\sqrt{2}} \left(a_i - a_i^\dagger\right).}
\end{array}
\end{equation}
Note that both $P_{i-1, i}$ and $S_{i, i+1}$ are unitary, both involves operators from two sites and $[P_{i-1, i}, P_{i, i+1}] \neq 0$, $[S_{i-1, i}, S_{i, i+1}] \neq 0$. $S_{i,i+1}$ only involves qutrit operators for one site, while $\left( a_i-a_i^\dagger \right)P_{i-1,i}$ only involves photonic operators from one site.

$U_\gamma$ acts like a displacement operator on the photonic variables, i.e.
\begin{equation}\label{polaronboson}
U_\gamma a_i U^\dagger_\gamma= a_i + \gamma \left[ a_i - a_i^\dagger, a_i \right] P_{i-1, i} = a_i+\gamma P_{i-1, i},
\end{equation}
which allows us to calculate the contribution to $H_\text{eff}$ from the photonic part of $H_B$ immediately as
\begin{equation}\label{photonic_bare}
\sum_i \langle \vec{\alpha} | U_\gamma a_i^\dagger a_i U^\dagger_\gamma | \vec{\alpha} \rangle = 
\sum_i \left( \alpha_i^2 + 2 \gamma \alpha_i P_{i-1, i}  + \gamma^2 P_{i-1,i}^2 \right).
\end{equation}

The transformation of the qutrit part of $H_B$ is more involved. Using the Hadamard lemma~\cite{shankar}, the transformed qutrit operators can be expanded as power series of nested commutators as
\begin{align}
U_\gamma \lambda^{(\alpha)}_i U_\gamma^\dagger = 
\lambda^{(\alpha)}_i - i \gamma \left[ S_{i,i+1}, \lambda^{(\alpha)}_i \right] + \nonumber \\
\frac{\left( -i \gamma \right)^2}{2!} \left[ S_{i,i+1}, \left[S_{i,i+1}, \lambda^{(\alpha)}_i \right] \right] + \dots.
\end{align} 
To compute coherent state expectation values, we use the formula (assuming $\alpha \in \mathbb{R}$):
\begin{align}\label{pn_moment}
\langle \alpha | p^n | \alpha \rangle = \begin{cases}
0 \,\,\,\,\,\,\,\,\,\,\,\,\,\,\,\,\,\,\,\,\,\,\,\,\,\,\,\,\,\,\,\, \text{ $n$ odd} \\
\Gamma(\frac{n+1}{2})/\sqrt{\pi} \,\,\,\,\, \text{ $n$ even} \\
\end{cases}
\end{align}
Using~(\ref{pn_moment}), we find
\begin{equation}
\begin{array}{l}
\displaystyle{- s \sum_i \langle \vec{\alpha} | U_\gamma \lambda^{(1)}_i U^\dagger_\gamma | \vec{\alpha} \rangle = - s f_1(\gamma) \sum_i \lambda^{(1)}_i}  \\ \\
\displaystyle{- \frac{\Omega}{\sqrt{3}} \sum_i \langle \vec{\alpha} | U_\gamma \lambda^{(8)}_i U^\dagger_\gamma | \vec{\alpha} \rangle = - \frac{\Omega}{\sqrt{3}} f_8(\gamma) \sum_i \lambda^{(8)}_i,}
\end{array}
\end{equation} 
where the functions of $\gamma$ can be expanded as rapidly converging power series
\begin{equation}
\begin{array}{l}
\displaystyle{f_1(\gamma) = 1-\gamma ^2+\frac{5 \gamma ^4}{6}-\frac{17 \gamma ^6}{30}+O\left(\gamma ^{8}\right)}\\ \\
\displaystyle{f_8 (\gamma) = 1-3 \gamma ^2+4 \gamma ^4-\frac{16 \gamma ^6}{5}+O\left(\gamma ^{8}\right).}
\end{array}
\end{equation}

It remains to calculate the contribution from the interaction Hamiltonian,
\begin{equation}
\begin{array}{l}
\displaystyle{g \sum_i \langle \vec{\alpha} | U_\gamma \left( a_i + a_i^\dagger \right) U_\gamma^\dagger U_\gamma P_{i-1,i}U^\dagger_\gamma | \vec{\alpha} \rangle =}\\ \\
\displaystyle{g \sum_i \langle \vec{\alpha} | \left( a_i + a_i^\dagger \right) U_\gamma  P_{i-1,i} U^\dagger_\gamma | \vec{\alpha} \rangle}\\ \\
\displaystyle{+2 g \gamma \sum_i P_{i-1, i} \langle \vec{\alpha} | U_\gamma  P_{i-1,i} U^\dagger_\gamma | \vec{\alpha} \rangle.}
\end{array}
\end{equation}
In the calculation, we will need to use the formula
\begin{align}\label{xpn_moment}
\langle \alpha | (a+a^\dagger) p^n | \alpha \rangle = \begin{cases}
in/\sqrt{2} \langle \alpha | p^{n-1} | \alpha \rangle \,\,\,\,\, \text{ $n$ odd} \\
2 \alpha \langle \alpha | p^{n} | \alpha \rangle \,\,\,\,\,\,\, \,\,\,\,\, \,\,\,\,\, \,\,\,\, \text{ $n$ even.} \\
\end{cases}
\end{align} 
Using~(\ref{xpn_moment}), we find
\begin{equation}\label{cross-term-2}
\begin{array}{l}
\displaystyle{g \sum_i \langle \vec{\alpha} | \left( a_i + a_i^\dagger \right) U_\gamma  P_{i-1,i} U^\dagger_\gamma | \vec{\alpha} \rangle =}\\ \\
\displaystyle{g \sum_\text{$i$ odd} \sum_{\beta=2,4,6} \left( h_{i-1, i}^{4\beta} \lambda^\beta_{i-1} + h_{i, i+1}^{4\beta} \lambda^\beta_{i} \right)}\\ \\
\displaystyle{+g \sum_\text{$i$ even} \sum_{\beta=2,4,6} \left( h_{i-1, i}^{6\beta} \lambda^\beta_{i-1} + h_{i, i+1}^{6\beta} \lambda^\beta_{i} \right) =}\\ \\
\displaystyle{g f_g(\gamma) \sum_i 2 \alpha_i P_{i-1, i}}
\end{array}
\end{equation}
since $h^{64}_{i-1,i} = h^{46}_{i-1,i} = 0$ for all $i$, $h^{42}_{i,i+1} = h^{62}_{i-1,i} = 0$ and $h^{42}_{i,i+1} = -h^{62}_{i-1,i} $ for $i$ odd and vice versa for $i$ even, while $h^{66}_{i,i+1} = h^{44}_{i, i+1} = f_g(\gamma)$ for all $i$, where 
\begin{equation}
f_g(\gamma) = 1-\frac{\gamma ^2}{2}+\frac{\gamma ^4}{6}-\frac{\gamma ^6}{30}+\frac{\gamma ^8}{210}+O\left(\gamma ^{10}\right).
\end{equation}
Note that the denominators are the primorial numbers~\cite{primorial}, so $f_g(\gamma)$ is the `primorial version' of $e^{-\gamma^2}$.

Furthermore, from~(\ref{pn_moment}), we know that only coherent state expectation values of even powers of $p^n$ are non-zero. By expanding in a power series using the Hadamard lemma and setting all terms which are odd in $p_{i-1}$ or $p_i$ to zero, we find
\begin{equation}
2 g \gamma \sum_i Q_{i-1, i} \langle \vec{\alpha} | U_\gamma  P_{i-1,i} U^\dagger_\gamma | \vec{\alpha} \rangle =
2 \gamma f_g(\gamma) \sum_i P_{i-1,i}^2.
\end{equation}

The expression~(\ref{cross-term-2}) and the cross-term in~(\ref{photonic_bare}) are the only terms which couple the photonic mean-field and qutrit degrees of freedom. Both terms can be eliminated by choosing $\gamma$ to be a root of the equation
\begin{equation}
g f_g(\gamma) = -\omega \gamma,
\end{equation}
which has a solution $\gamma \sim -g/\omega$ close to $g/\omega = 0$. By this choice, the photonic and qutrit degrees of freedom are completely decoupled. Once decoupled, the photonic degrees of freedom can be minimized independently to $\alpha_i = 0$. In this sense, the polaron ansatz is a ``dynamic displacement'' of the photonic vacuum. The ansatz neglects any contribution to quantum fluctuations from the resonators around this vacuum state.

We found that the photonic part of the effective qutrit Hamiltonian was minimized to zero, so the remaining part is a qutrit Hamiltonian:
\begin{equation}
\begin{array}{lll}
H_\text{eff} & = & \displaystyle{ - s f_1(\gamma) \sum_i \lambda^{(1)}_i - \frac{\Omega}{\sqrt{3}} f_8(\gamma) \sum_i \lambda^{(8)}_i}\\ \\
& & \displaystyle{+\Big( \omega \gamma^2 + 2 \gamma f_g(\gamma)  \Big) \sum_i P_{i-1,i}^2.}
\end{array}
\end{equation}
Assuming periodic boundary conditions, we find (the constant term is dropped):
\begin{equation}
\begin{array}{lll}
\displaystyle{\sum_i P_{i-1,i}^2} & = & \displaystyle{\sum_\text{$i$ odd} \Big( (\lambda^{(4)}_{i-1})^2 + (\lambda^{(4)}_{i-1})^2 + 2 \lambda^{(4)}_{i-1} \lambda^{(4)}_i \Big)}\\ \\
& & \displaystyle{+\sum_\text{$i$ even} \Big( (\lambda^{(6)}_{i-1})^2 + (\lambda^{(6)}_{i-1})^2 + 2 \lambda^{(6)}_{i-1} \lambda^{(6)}_i \Big)}\\ \\
& = & \displaystyle{2 \sum_\text{$i$ odd} \lambda^{(4)}_{i-1} \lambda^{(4)}_i + 2 \sum_\text{$i$ even} \lambda^{(6)}_{i-1} \lambda^{(6)}_i}\\ \\
& & \displaystyle{+\sum_i \left( \frac{4}{3} \mathbb{I}_{3\times3} - \frac{1}{\sqrt{3}} \lambda^{(8)}_i \right),}
\end{array}
\end{equation}
such that the effective qutrit Hamiltonian can be written as
\begin{equation}
\begin{array}{lll}
H_\text{eff} & = & \displaystyle{ -\tilde{s} \sum_i \lambda^{(1)}_i - \frac{\tilde{\Omega}}{\sqrt{3}} \sum_i \lambda^{(8)}_i - J \sum_\text{$i$ odd} \lambda^{(4)}_{i-1} \lambda^{(4)}_{i}}\\ \\
& & \displaystyle{ - J \sum_\text{$i$ even} \lambda^{(6)}_{i-1} \lambda^{(6)}_{i},}
\end{array}
\end{equation}
where
\begin{equation}
\begin{array}{l}
\tilde{s} = f_1(\gamma) s, \\ \\
\tilde{\Omega} = f_8(\gamma) \Omega + J/2, \\ \\
\displaystyle{J = -2 \Big( \omega \gamma^2 + 2 \gamma f_g(\gamma) \Big) = \frac{2-g}{g} 2 \omega \gamma^2.} 
\end{array}
\end{equation}


\end{document}